\documentclass[useAMS,usenatbib]{mn2e}
\usepackage{epsfig}

\def\het{{\it HETE-2}}
\def\sax{{\it BeppoSAX}}
\def\sw{{\it Swift}}
\def\dof{{\rm dof}}

\title[Correlation between Variability and Luminosity of
GRBs]{Improved Correlation between the Variability and Peak Luminosity of
Gamma-Ray Bursts}
\author[Li-Xin Li and Bohdan Paczy\'nski]{Li-Xin Li$^{1}$\thanks{E-mail:
lxl@mpa-garching.mpg.de; bp@astro.princetion.edu} and Bohdan
Paczy\'nski$^{2}$\footnotemark[1]\\
$^{1}$Max-Planck-Institut f\"ur Astrophysik, 85741 Garching, Germany\\
$^{2}$Princeton University Observatory, Princeton, NJ 08544, USA}

\begin{document}

\date{}


\pagerange{\pageref{firstpage}--\pageref{lastpage}} \pubyear{2005}

\maketitle

\label{firstpage}

\begin{abstract}
A new procedure for smoothing a gamma-ray burst (GRB) lightcurve and 
calculating its variability is presented. Applying the procedure to a 
sample of 25 long GRBs, we have obtained a very tight correlation between 
the variability and the peak luminosity. The only significant outlier in 
the sample is GRB 030329. With this outlier excluded, the data scatter is
reduced by a factor of $\sim 3$ compared to that of Guidorzi et al. (2005),
measured by the deviation of fit. Possible causes for the outlier are 
discussed.
\end{abstract}


\section{Introduction}

Despite many exciting progresses in observations, the nature of Gamma-Ray 
Bursts (GRBs) remains to be a big puzzle [see \citet{par00} and \citet{pir04} 
for recent reviews]. In such a situation, it is extremely important to 
identify some good correlations between the apparent (easy to measure or
calculate) and 
intrinsic properties of GRBs. Several such correlations have indeed been 
found. For example, an anti-correlation between the the peak luminosity and 
spectral lag of GRBs has been found by \citet{nor00}, and a correlation 
between the peak luminosity and the variability of GRB lightcurves has been 
found by Fenimore \& Ramirez-Ruiz (2000, hereafter FR00) and Reichart et al. 
(2001, hereafter R01). A correlation between the total isotropic energy and 
the peak energy of the spectrum \citep{ama02}, or the collimation-corrected 
total energy and the peak energy of the spectrum \citep{ghi04}, has also been 
discovered.

Recently, Guidorzi et al. (2005, hereafter G05) tested the correlation between 
the variability and peak luminosity of GRBs, using an expanded sample of 32 
GRBs with measured redshifts. The definitions of the variability and the peak 
luminosity are the same, but the size of the GRB sample of G05 is about three 
times bigger than that of R01. The existence of a correlation was confirmed, 
but the scatter in the correlation is significantly larger than that found by 
R01 (see, however, Reichart \& Nysewander 2005). Although the issue is in 
debate \citep{gui05b,rei05,rei05b}, it is clear that with the definition of 
variability given by R01, the correlation between the variability and the 
peak luminosity is not tight. 

In this paper, we present a new definition for the variability of GRB 
lightcurves. We then apply it to a sample of 25 long duration GRBs with measured
redshifts, whose data are publicly available. We show that, with the new
definition of the variability, the correlation between the variability and
the peak luminosity of GRBs is significantly improved: the data scatter 
is considerably reduced.

\section{New Definition of the Variability}

To measure the variability of a lightcurve, first we must define a 
{\it reference lightcurve}. The reference lightcurve should be sufficiently 
smoother than the original raw lightcurve. Since an ultimate model for 
GRBs does not exist yet, there is no first principle guiding us in 
choosing a reference lightcurve. What people usually do is to smooth
the raw lightcurve with a linear ``box car'' filter (or moving window), 
which smoothes the lightcurve with linear average (FR00; R01). 

Here we use a {\it Savitzky-Golay filter} \citep{pre02} to smooth a 
lightcurve. The Savitzky-Golay filter is a more general and more powerful 
approach for smoothing noisy data than the linear box-car filter. The basic 
idea of Savitzky-Golay filtering is to approximate the underlying function 
(i.e., the reference lightcurve) within the moving window by a polynomial 
of higher order. An advantage of the Savitzky-Golay filter to the linear 
filter is that the former preserves high moments while the latter does not. 

A Savitzky-Golay filter is specified by three numbers: the order
of the polynomial ($m$), the number of points used to the left of a data
point ($n_{\rm L}$), and the number of points used to the right of a data
point ($n_{\rm R}$). To apply the Savitzky-Golay filter, the data must be
binned with constant spacing. For more details see \citet{pre02}.

We use a third order Savitzky-Golay filter. That is, we set $m=3$. The width 
of the filter, i.e. $n_{\rm P}\equiv n_{\rm L} + n_{\rm R}+ 1$, is determined 
with the approach of R01: $n_{\rm P}$ is set to be equal to the number of 
data points corresponding to a timescale $T_f$--the time spanned by the 
brightest $100f\%$ of the total counts above the background (for details see 
R01). It turns out that $f=0.5$ most suits our purpose (rather than the $f=
0.45$ used by R01 and G05; see Section~5). So, throughout the paper we use 
$n_{\rm P}$ determined by $T_{0.5}$. 

We then define $n_{\rm L}={\rm int} [(n_{\rm P}-1)/2]$, and $n_{\rm R} = 
n_{\rm P}-n_{\rm L}-1$. If $n_{\rm P}$ is odd, we have $n_{\rm R}=n_{\rm L}$, 
i.e. the filter is symmetric about the point to smooth. If $n_{\rm P}$ is even, 
we have $n_{\rm R}=n_{\rm L}+1$, the filter is asymmetric. In this case, we 
smooth the lightcurve twice: first we use the $n_{\rm L}$ and $n_{\rm R}$ 
defined above, then we switch $n_{\rm L}$ and $n_{\rm R}$. The results are 
then averaged.

Suppose we have obtained a reference lightcurve by applying the Savitzky-Golay 
filter to the raw lightcurve. Let us denote the count of the raw data in the 
$i$-th time bin by $C_i$, the count given by the reference lightcurve by $Y_i$. 
The total squared deviation of the raw lightcurve from the reference lightcurve 
is then $\sum_{i=1}^{N_{\rm bin}} \left(C_i -Y_i\right)^2$, where $N_{\rm 
bin}$ is the total number of bins to be summed. We obtain the intrinsic squared 
deviation by subtracting the Poisson noise $N_{\rm Poisson}$
\begin{eqnarray}
	\Delta C^2 = W \sum_{i=1}^{N_{\rm bin}} \left(C_i -Y_i\right)^2
	     	- N_{\rm Poisson} \;,
	\label{dc2}
\end{eqnarray}
where $W \equiv n_{\rm P}/\left(n_{\rm P}-m-1\right)$ is a statistical weight
accounting for the fact that among the $n_{\rm P}$ data points only
$n_{\rm P}-m-1$ are statistically independent. The inclusion of $W$ allows
us to apply the variability definition to any lightcurve with $n_{\rm P}
>m+1$.

The summation in equation~(\ref{dc2}) is from time $t_1$ (corresponding to 
$i=1$) to time $t_2$ (corresponding to $i=N_{\rm bin}$), enclosing a major 
part of the lightcurve. Following FR00 and R01, we define $t_1$ to be the 
start of $T_{90}$, $t_2$ to be the end of $T_{90}$, where $T_{90}$ is the 
time during which the cumulative counts of the GRB increase from 5\% to 
95\% above background \citep{kou93}. 

The Poisson noise is calculated by
\begin{eqnarray}
	N_{\rm Poisson} = \sum_{i=1}^{N_{\rm bin}} \left[C_i + (\xi-1) 
		C_{{\rm bg},i}\right] \;,
\end{eqnarray}
where $C_{{\rm bg},i}$ is the background, the factor $\xi$ is the ratio of
the background fluctuation to the Poisson noise of the background (given by
the reduced $\chi^2$ of the background fit). For GRBs detected by \sw\ we 
found that $\xi$ is often significantly larger than unity, indicating that 
the background fluctuation is quite non-Poissonian. 

The {\it variability} of the lightcurve is defined by the normalized squared 
deviation. We find that the following definition leads to the tightest 
correlation between the variability and the peak luminosity
\begin{eqnarray}
	V = \frac{\Delta C^2}{\left(N_{\rm bin}-1\right) 
		C_{\rm max}^2} \;, \label{v}
\end{eqnarray}
where $\Delta C^2/\left(N_{\rm bin}-1\right)$ is the average of the squared 
deviation, $C_{\rm max}$ is the net peak count (i.e., the background is 
subtracted).

Our variability is defined in the observer's frame, so that the information
of GRB redshift is not needed. This not only makes the computation of 
variability simple, but also eliminates an uncertainty arising from the 
assumption about the dependence of lightcurve variability on photon energy.
The effect of GRB redshift was considered by R01 and FR00 who defined their 
variabilities in the GRB frame, but the dependence of variability on 
redshift turned out to be very weak. 

As we applied our smoothing procedure to the GRBs in the sample described in
the next section, we found that the tightest correlation between the variability
and the peak luminosity is obtained if we iteratively apply the Savitzky-Golay 
filter $N_{\rm iter}$ times, where $N_{\rm iter}$ is the integer closest to 
$T_{90}/T_f$ (i.e., $N_{\rm iter}$ is roughly the number of moving windows 
contained in $T_{90}$).

In summary, our definition of variability differs from that of FR00 and R01 in 
the following aspects: (1) Our variability is defined in the observer's frame, 
while the variabilities of FR00 and R01 are defined in the GRB's frame. 
(2) We normalize the average of the squared deviation by the squared peak 
count (the same as FR00), while R01 normalize the total squared deviation by 
the sum of squared counts. (3) FR00 and R01 use a linear box-car filter, 
while we use a nonlinear Savitzky-Golay filter.

\begin{table*}
\centering
\begin{minipage}{140mm}
\caption{Variability versus peak luminosity for 25 GRBs with known redshift.}
\label{grb}
\begin{tabular}{lrlrrrr}
\hline
GRB &  $z^{\rm (a)}$ & ~~Mission$^{\rm (b)}$ & ~$N_{\rm iter}^{\rm (c)}$~~ & $V^{\rm (d)}$ & $L^{\rm (e)}$ & $z$ Refs.$^{\rm (f)}$\\
\hline
970508  & 0.835  & ~~B/BS/U/K     & 6~~ & $0.0018 \pm 0.0009$   & $ 9.43  \pm   1.89$ & 1\\
971214  & 3.418  & ~~B/BS/U/K/N/R & 4~~ & $0.0106  \pm 0.0018$  & $ 360   \pm     65$ & 2\\
980425  & 0.0085 & ~~B/BS/U/K     & 6~~ & $0.00041\pm 0.00041$  & $ 0.0007\pm 0.0002$ & 3\\
980703  & 0.966  & ~~B/BS/U/K/R   & 4~~ & $0.0030 \pm 0.0005$   & $ 26.4  \pm    5.6$ & 4\\
990123  & 1.6004 & ~~B/BS/U/K     & 3~~ & $0.0100 \pm 0.0017$   & $ 840   \pm    121$ & 5\\
990506  & 1.3    & ~~B/BS/U/K/R   & 8~~ & $0.0083 \pm 0.0004$   & $ 583   \pm    121$ & 6\\
990510  & 1.619  & ~~B/BS/U/K/N   & 12~~ & $0.0064 \pm 0.0003$   & $ 300   \pm     50$ & 7\\
991216  & 1.02   & ~~B/BS/U/N     & 4~~ & $0.0132 \pm 0.0004$   & $ 1398  \pm    200$ & 8\\
000131  & 4.5    & ~~B/U/K/N      & 8~~ & $0.0113 \pm 0.0007$   & $ 3600  \pm    900$ & 9\\
010921  & 0.45   & ~~H/BS/U/K     & 3~~ & $0.0020 \pm 0.0017$   & $ 8.0   \pm    2.0$ & 10\\
020124  & 3.198  & ~~H/U/K        & 5~~ & $0.0097 \pm 0.0045$   & $ 300   \pm     60$ & 11\\
020813  & 1.25   & ~~H/U/K/O      & 5~~ & $0.0100 \pm 0.0013$   & $ 340   \pm     70$ & 12\\
030115a & 2.2    & ~~H            & 12~~ & $0.0057 \pm 0.0036$   & $ 57.0  \pm    8.0$ & 13\\
030328  & 1.52   & ~~H/U/K        & 3~~ & $0.0043 \pm 0.0015$   & $ 90    \pm     18$ & 14\\
030329  & 0.168  & ~~H/U/K/O/RH   & 4~~ & $0.0070 \pm 0.0010$   & $ 6.1   \pm    1.2$ & 15\\
030528  & 0.782  & ~~H            & 4~~ & $0.0018 \pm 0.0023$   & $ 1.4   \pm    0.5$ & 16\\
041006  & 0.712  & ~~H/K/RH       & 6~~ & $0.0037 \pm 0.0008$   & $ 66    \pm     10$ & 17\\
050401  & 2.90   & ~~BSw          & 7~~ & $0.0120 \pm 0.0022$   & $ 740   \pm    100$ & 18\\
050408  & 1.2357 & ~~H            & 9~~ & $0.0072 \pm 0.0033$   & $ 148   \pm     67$ & 19\\
050505  & 4.27   & ~~BSw          & 8~~ & $0.0090 \pm 0.0051$   & $ 250   \pm     50$ & 20\\
050525  & 0.606  & ~~BSw          & 3~~ & $0.0071 \pm 0.0012$   & $ 80    \pm     10$ & 21\\
050603  & 2.821  & ~~BSw          & 6~~ & $0.0100 \pm 0.0006$   & $ 1200  \pm    200$ & 22\\
050802  & 1.71   & ~~BSw          & 3~~ & $0.0042 \pm 0.0038$   & $ 104   \pm     23$ & 23\\
050803  & 0.422  & ~~BSw          & 6~~ & $0.0026 \pm 0.0009$   & $ 2.4   \pm    0.6$ & 24\\
050820a & 2.612  & ~~BSw          & 5~~ & $0.0058 \pm 0.0038$   & $ 63    \pm     13$ & 25\\
\hline
\end{tabular}
\begin{list}{}{}
\item[$^{\rm (a)}$] Measured redshift.
\item[$^{\rm (b)}$] Mission: B (BATSE/{\it CGRO}), H ({\it HETE-2}), BS (\sax), 
K (Konus/{\it WIND}), BSw (BAT/{\it Swift}), U ({\it Ulysses}), S ({\it
SROSS-C}), N ({\it NEAR}), R ({\it RossiXTE}), O ({\it Mars Odyssey}), RH ({\it 
RHESSI}): the data used are taken from the first mission mentioned.
\item[$^{\rm (c)}$] Number of iterations in applying the Savitzky-Golay filter
(see Section~2).
\item[$^{\rm (d)}$] Calculated variability (equation~\ref{v}) and error.
\item[$^{\rm (e)}$] Isotropic-equivalent peak luminosity in $10^{50}$~erg s$^{-1}$
in the rest-frame 100--1000~keV band, for peak fluxes measured on a 1-s 
time-scale ($H_0 = 65$ km s$^{-1}$ Mpc$^{-1}$, $\Omega_m = 0.3$, and 
$\Omega_{\Lambda} = 0.7$).
\item[$^{\rm (f)}$] References for the redshift measurements:
(1) \citet{Metzger97}, (2) \citet{Kulkarni98}, (3) \citet{Tinney98},
(4) \citet{Djorgovski98}, (5) \citet{Kulkarni99}, (6) \citet{Bloom03}, 
(7) \citet{Beuermann99}, (8) \citet{Vreeswijk99}, (9) \citet{Andersen00}, 
(10) \citet{Djorgovski01b}, (11) \citet{Hjorth03}, (12) \citet{Price02b}, 
(13) http://space.mit.edu/HETE/Bursts/Data/ and \citet{smith05}, 
(14) \citet{Martini03}, (15) \citet{Greiner03b}, (16) \citet{rau05},
(17) \citet{Fugazza04}, (18) \citet{Fynbo05a}, (19) \citet{Berger05a} and
\citet{prochaska05a}, (20) \citet{Berger05_b}, (21) \citet{Foley05}, 
(22) \citet{Berger05_c}, (23) \citet{Fynbo05b}, (24) \citet{bloom05}, 
(25) \citet{prochaska05b} and \citet{ledoux05}.
\end{list}
\end{minipage}
\end{table*}

\begin{figure*}
\vspace{2pt}
\includegraphics[angle=0,scale=0.8]{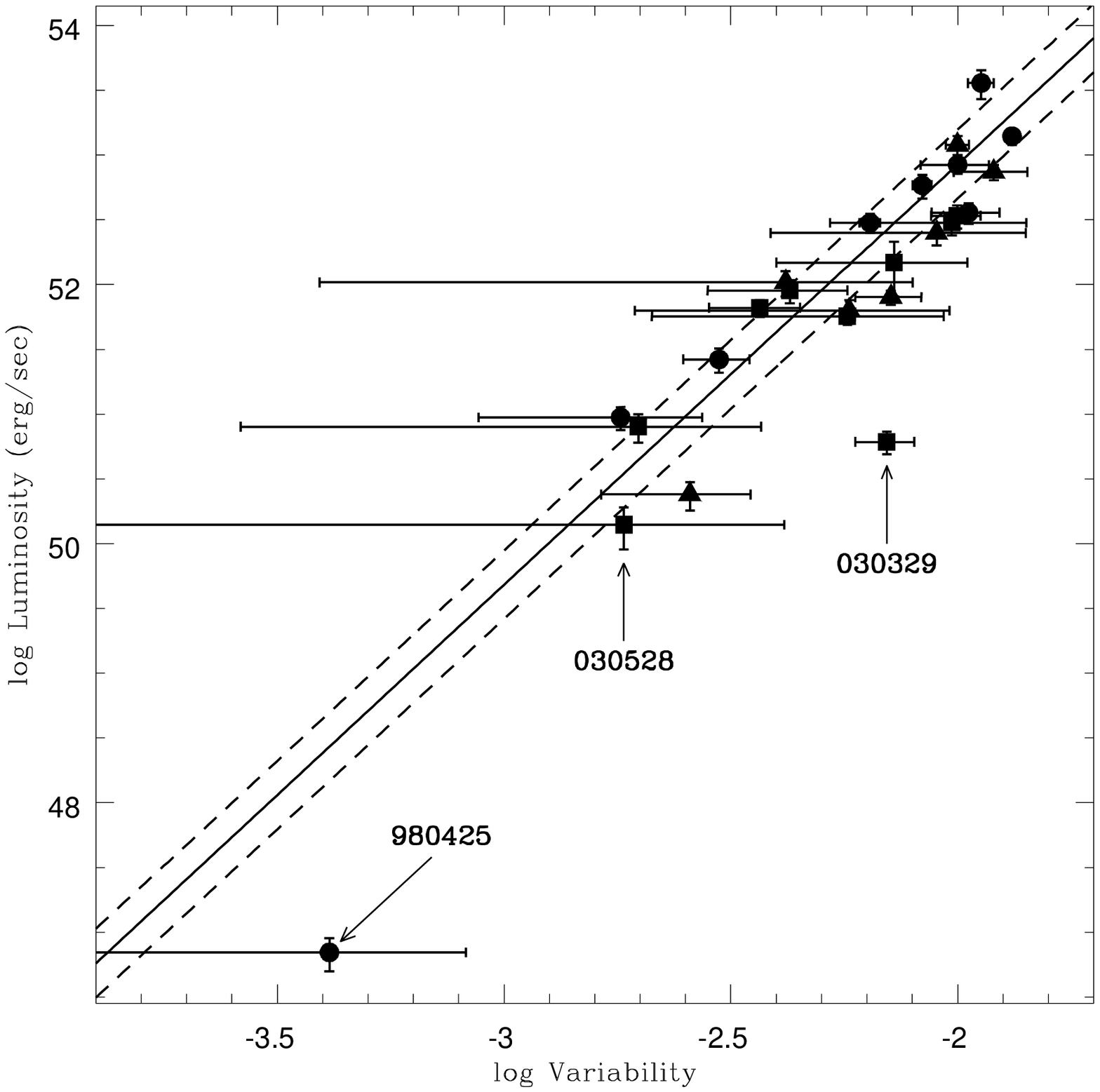}
\caption{Peak luminosity versus variability of 25 GRBs. The solid straight line 
is the least-$\chi^2$ linear fit to the data, with GRBs 980425, 030329, and 
030528 excluded (see the text): $\log L = 3.25 \log V + 59.42$ ($L$ in units 
of erg/sec). The $\chi^2/\dof=38.58/20$=1.93. The two dashed lines mark the
1-$\sigma$ deviation of the fit. (Filled circles: GRBs detected by BATSE; 
Squares: GRBs detected by \het; Triangles: GRBs detected by BAT/\sw.)
}
\label{fig1}
\end{figure*}

\section{Description of the Sample}

Our GRB sample contains 19 GRBs from the sample in G05, and 6 more GRBs detected 
by FREGATE/\het\ and BAT/\sw. So, the total number of GRBs in our sample is 25. 
They are listed in Table~\ref{grb}, with measured redshift, calculated 
isotropic-equivalent peak luminosity, calculated variability, and the number
of iterations in applying the Savitzky-Golay filter. We have chosen 
to use the GRBs with data available publicly, which include GRBs detected by 
BATSE/{\it CGRO}, \het, and BAT/\sw.\footnote{Although the lightcurve data of 
some GRBs detected by Konus/{\it Wind} are available from 
http://gcn.gsfc.nasa.gov/gcn/ konus\_grbs.html, we have chosen not to use them 
since the data are highly incomplete in the part of pre- and post-burst.} 
To obtain a reliable calculation of variability, we have only selected GRBs 
with more than 30\% of total counts above the 3-$\sigma$ of background. 
As a result, those GRBs with too low signal-to-noise ratios are not included in 
our sample. 

The 19 GRBs from G05 are: 970508, 971214, 980425, 980703, 990123, 
990506, 990510, 991216, 000131, 010921, 020124, 020813, 030328, 030329, 041006, 
050401, 050505, 050525, and 050603. Their peak luminosities are taken from the 
same paper. The rest 13 GRBs in G05 are not included in our sample for various
reasons: either their data are not publicly available (noticeably the 7 GRBs
detected by \sax\ but not by BATSE or \het), or their data are incomplete
or have too low signal-to-noise ratios. Although 050315 and 050319 in G05 were
detected by BAT/\sw, their data were not available to us since as this paper 
was written the archive of \sw\ only contained data taken after 1 April 2005.

The 6 newly added GRBs are 030115a, 030528, 050408 (from \het), 050802, 050803, 
and 050820a (from \sw). Their peak luminosities are calculated in exactly the 
same way as in R01 and G05. For GRBs detected by \het, we used the spectral fit 
provided by \citet{sak04} and the {\it HETE} website. For GRBs detected by \sw, 
peak spectra are extracted from their event files, then fitted with a power law 
which is sufficient for all cases (in the energy range 15-350 keV). 

Although \citet{gol05}, \citet{pal05}, and \citet{cum05} have reported the 
detection of a second, larger episode of emission from GRB 050820a (by both
Konus/{\it Wind} and BAT/\sw), only the data for the first episode 
from \sw\ are available to us. Thus, for GRB 050820a, the variability and peak 
luminosity listed in Table~1 refer to its first episode (of duration $\sim 30$ 
sec).

The time bin of each GRB lightcurve is as follow. For GRBs detected by BATSE, 
except 000131, the time bin is 64-ms. GRB 000131 does not have 
a 64-ms lightcurve, so we use its 1.024-sec lightcurve. Although a bin of 
1.024-sec is somewhat large, the duration of 000131 is about 100-sec which is 
much larger than its time bin. For GRBs detected by \het, the time bin is 
164-ms. While for GRBs detected by BAT/\sw, we extract 64-ms lightcurves from 
their event files available in the public data archive.

As we analyze the GRB lightcurves we use the total counts in the energy
range specified as follow: 25-300 keV for GRBs detected by BATSE, 30-400 keV 
for those detected by \het, and 25-350 keV for those detected by BAT/\sw. 
We have tried to make the range of energy for GRBs detected by different 
instruments have largest overlap.

Finally, we remark that for the GRBs detected by BATSE, the lightcurve data
in the pre- and post-burst sections were divided equally into 64-ms time bins
from their original 1.024-sec time bins. When we calculate the Poisson noise 
in those sections, we multiply the result by a factor of $0.0625$ ($=0.064/1.024$) 
to take into account the reduction in the Poisson noise arising from the 
change in time bin.

\section{Results}

We have applied the smoothing procedure described in Section~2 to our GRB 
sample. The obtained variabilities and errors, as well as the number of 
iterations in applying the Savitzky-Golay filter, are listed in Table~\ref{grb}. 
The errors of variability are principally caused by photon noises. However, 
the uncertainties arising from changing $N_{\rm iter}$ (see Section~2) by 
$\pm 1$ are also taken into account: we find the changes in the variability 
for $N_{\rm iter}\pm 1$ and add in quadrature the maximum to the statistical 
error.

The peak luminosity versus variability is shown in Fig.~\ref{fig1}. Clearly a very
tight correlation between the two quantities exists, with only one prominent 
outlier: GRB 030329. Given the large error in its variability, it is unclear if 
GRB 980425---which is famous for its smallest redshift, lowest peak luminosity
and least total energy, and association with SN 1998bw \citep{gal98}---is off 
trend as in other kind of correlations (e.g., the total energy-peak energy 
correlation). In fact, the error in variability is larger than or about the 
same value as the variability itself for 980425 and 030528 (see Table~1). 

We have made a least-$\chi^2$ linear fit to $\log L - \log V$, where $L$ is the 
peak luminosity in erg/sec. To take into account both the errors in $L$ and $V$, 
we have made use of the program {\sf fitexy} in \citet{pre02}. The asymmetric
errors in $\log L$ and $\log V$ are averaged. That is, we take 
\begin{eqnarray}
	\sigma_{\log q} = \frac{1}{2}\left[\log \left(q+\sigma_q\right)
		- \log \left(q-\sigma_q\right)
		\right] \; ,
\end{eqnarray}
where $q = L$ or $V$. Since the errors in $\log V$ are infinite for 980425 and 
030528, these two GRBs do not contribute to the total $\chi^2$. GRB 030329 was 
not considered due to its large offset. With 980425, 030329, and 030528 
excluded, the total number of GRBs in the final sample to fit is 22. We have 
obtained the following result (the solid straight line in Fig.~\ref{fig1})
\begin{eqnarray}
	\log L = a \log V + b \;, \label{lllv}
\end{eqnarray}
where
\begin{eqnarray}
	a = 3.25 \pm 0.26 \;, \hspace{1cm}
	b = 59.42 \pm 0.53 \;. \label{ab}
\end{eqnarray}
The reduced chi-square $\chi_r^2 \equiv \chi^2/\dof = 1.93$, where the degree 
of freedom $\dof =22-2 =20$. The smallness of $\chi^2_r$ indicates a very 
tight correlation between the peak luminosity and variability, for details
see Section~5.

The two dashed lines in Fig.~\ref{fig1} mark the 1-$\sigma$ width of the
fit, measuring the scatter of data and defined by the deviation of fit 
(see Section~5). However, this should not be mixed with the error in the
predicted $\log L$, which we discuss below.

\begin{figure}
\vspace{2pt}
\includegraphics[angle=0,scale=0.47]{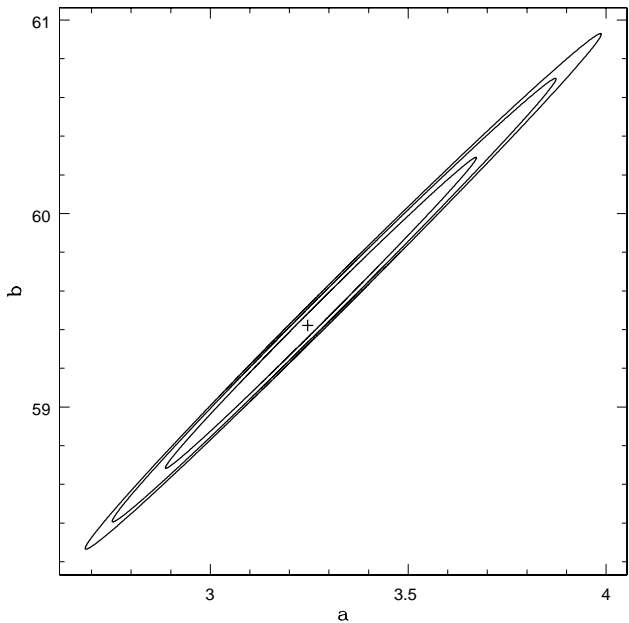}
\caption{The confidence intervals in the $a-b$ space. The cross sign denotes 
the values of $a$ and $b$ at the minimum of $\chi^2$ ($a=3.25$, $b=59.42$,
$\chi^2=38.58$). The contours correspond to the increase in $\chi^2$, $\Delta 
\chi^2 = 2.30$, $4.61$, and $6.17$ respectively. The region enclosed by each 
contour contains $68.3\%$, $90\%$, and $95.4\%$ of normally distributed data.
}
\label{fig2}
\end{figure} 

\begin{figure}
\vspace{2pt}
\includegraphics[angle=-90,scale=0.39]{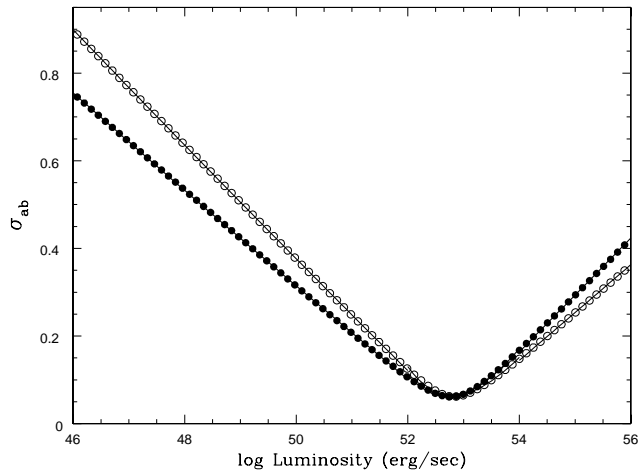}
\caption{The error in the $\log L$ predicted by equation~(\ref{lllv}), due to 
the uncertainties in $a$ and $b$. Filled circles are numerical solutions for 
the upward error ($\sigma_{ab,1}$). Open circles are numerical solutions for 
the downward error ($\sigma_{ab,2}$). Solid lines are analytical 
approximations to the numerical solutions given by 
equations~(\ref{eq1})--(\ref{eq6}).
}
\label{fig3}
\end{figure}

In Fig.~\ref{fig2} we plot the confidence intervals in the $a-b$ plane. The
best fit values of $a$ and $b$, which correspond to the minimum of $\chi^2$,
are indicated by the cross sign in the figure. The three contours enclose the
regions containing, respectively, $68.3\%$, $90\%$, and $95.4\%$ of normally
distributed data. The highly elongated shape of the contours indicates that 
$a$ and $b$ are highly correlated.

For any measured value of $\log V$, equation~(\ref{lllv}) predicts a value of 
$\log L$. The error in the predicted $\log L$ comes from two sources: the
error in $\log V$, and the uncertainties in $a$ and $b$
\begin{eqnarray}
	\sigma_{\log L} = \sqrt{a^2 \sigma_{\log V}^2 + \sigma_{ab}^2} \;.
	\label{sigma}
\end{eqnarray}

Generally, the errors are asymmetric in the upward and downward directions. 
Let us denote by $\sigma_{ab,1}$ the upward error due to uncertainties in $a$ 
and $b$, by $\sigma_{ab,2}$ the downward error. That is, the predicted log 
luminosity is $\log L_{-\sigma_{ab,2}}^{+\sigma_{ab,1}}$, if $\sigma_{\log V} 
= 0$. In Fig.~\ref{fig3} we show the numerically calculated values of 
$\sigma_{ab,1}$ (filled circles) and $\sigma_{ab,2}$ (open circles) versus 
the predicted log luminosity. From Fig.~\ref{fig2} we find that $\delta b 
\sim 2.05\, \delta a$ when $\delta a$ and $\delta b$ are not close to zero, 
which leads to $\sigma_{ab}\sim (\log V +2.05) \delta a$. Thus, $\sigma_{ab}$ 
is minimized at $\log V \approx -2.05$, i.e. at $\log L \approx 52.8$, 
consistent with the numerical results.

For the convenience of application of the $\log L-\log V$ 
relation~(\ref{lllv}) to the prediction of GRB's peak luminosity, we have 
attempted to approximate the numerical results for $\sigma_{ab}$ by several 
analytic formulae. The results are summarized in Appendix~A, and represented 
by the solid lines in Fig.~\ref{fig3}.

Now we apply the above results to GRB 030329 to check how far it deviates from 
the $\log L-\log V$ relation. From Table~1, for GRB 030329 we have $V =0.0070$ 
and $\sigma_V = 0.0010$. Submitting $\log V = -2.15$ into equation~(\ref{lllv}), 
we obtain a predicted log luminosity for GRB 030329: $\log L = 52.4$. Since 
GRB 030329 lies below the straight line in Fig.~\ref{fig1}, the relevant error 
in the predicted log luminosity is determined by $\sigma_{ab,2}$ and 
$\sigma_{\log V} = \log V - \log \left(V-\sigma_V\right) \approx 0.067$. By 
equation~(\ref{eq4}) we have $\sigma_{ab,2} \approx 0.08$. Then, by 
equation~(\ref{sigma}), the total error in the predicted luminosity is 
$\sigma_{\log L} \approx 0.23$. From Table~1, the measured luminosity of 
GRB 030329 is $\log L \approx 50.8$, with an upward measured error 
$\sigma_{\log L} \approx 0.08$ that is smaller than the error in the 
predicted log luminosity. So, GRB 030329 deviates from the correlation 
(the straight line in Fig.~\ref{fig1}) by $(52.4-50.8)/0.23 \approx 7\,\sigma$.

\section{Comparison to Guidorzi et al.'s Results}

To make a quantitative comparison of our results to that of G05, we need a 
parameter that measures the scatter of data points around the best fit model. 
The parameter should not be sensitive to the variation in data errors caused 
by different definitions of variability. \citet{rei01a} has provided such a 
parameter based on the maximum-likelihood method, which is called the ``sample 
variance'', or ``slop''. However, here we choose to use the ``variance of 
fit'' defined with the approach of least-$\chi^2$ \citep{bev92}, which is 
much less sophisticated but enough for our current purpose.

Suppose a data set $\{x_i,y_i\}$, with errors $\{\sigma_{x,i},\sigma_{y,i}\}$,
is fitted by a model $y=f(x)$. The fit minimizes the $\chi^2$ defined by
\begin{eqnarray}
	\chi^2 \equiv \sum_{i=1}^{N}\frac{1}{\sigma_i^2}\left[y_i-
		f\left(x_i\right)\right]^2 \;,
	\label{chi2}
\end{eqnarray}
where $N$ is the total number of data points, and
\begin{eqnarray}
	\sigma_i^2 = \sigma_{y,i}^2 + \left(\frac{df}{dx}\right)_{x=x_i}^2
		\sigma_{x,i}^2 \;.
\end{eqnarray}
The reduced $\chi^2$, i.e., the $\chi^2$ per degree of freedom, is
\begin{eqnarray}
	\chi_r^2 = \chi^2/\dof = \chi^2/(N-p) \;,
\end{eqnarray}
where $p$ is the number of parameters in the function $f(x)$. 

\begin{figure}
\vspace{2pt}
\includegraphics[angle=0,scale=0.48]{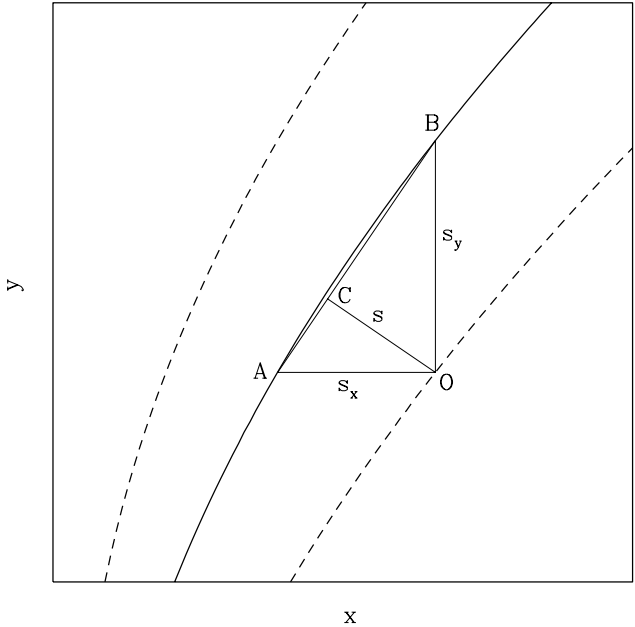}
\caption{Definition of the variance of fit. The thick solid curve is the
fit $y=f(x)$. The $y$-component of the variance of fit, $s_y^2$, is defined 
by equation~(\ref{sfit1}). The two dashed curves, called ``the 1-$\sigma$ 
width'' of $y=f(x)$, are defined by $y=f(x)\pm s_y$. The $x$-component of 
the variance of fit, $s_x^2$, is related to $s_y^2$ by equation~(\ref{sfit2}).
The scatter of data points around $y=f(x)$ is measured by $s=s_xs_y/\sqrt{
s_x^2+s_y^2}$, the length of the straight line $OC$ that is perpendicular 
to the straight line $AB$. [The straight line $AB$ locally approximates the 
curve $y=f(x)$ when $s_y$ is sufficiently small or the curvature of the
curve is negligible.]
}
\label{fig4}
\end{figure}

The variance of the fit, measured along the $y$-direction, is defined by 
\citep{bev92}
\begin{eqnarray}
	s_y^2 \equiv \frac{\langle\sigma_i^2\rangle}{N-p}\sum_{i=1}^N
		\frac{1}{\sigma_i^2}\left[y_i-f\left(x_i\right)\right]^2 \;,
	\label{sfit1}
\end{eqnarray}
where $\langle\sigma_i^2\rangle$ is the weighted average of the individual 
data variance
\begin{eqnarray}
	\langle\sigma_i^2\rangle \equiv \left[\frac{1}{N}\sum_{i=1}^N
		\frac{1}{\sigma_i^2}\right]^{-1} \;.
\end{eqnarray}
The {\em variance of fit} measures the spread of data around the model. 
We denote the variance of fit by $s^2$ (then $s$ is the {\em deviation of fit}), 
to distinguish it from the {\em data variance} $\sigma^2$ which represents the 
error in the measurement of each data point.

The appearance of $\langle\sigma_i^2\rangle$ in the numerator and $\sigma_i^2$
in the denominator in equation~(\ref{sfit1}) results that the variance of
fit is not sensitive to the variation in data variance. 

By the definition of $\chi_r^2$, equation~(\ref{sfit1}) can be recast into
a simpler form
\begin{eqnarray}
	s_y^2 = \langle \sigma_i^2 \rangle \chi_r^2\;.
	\label{sfit}
\end{eqnarray}

We define the variance of the fit measured along the $x$-direction by
\begin{eqnarray}
	s_x^2 = s_y^2\left(\frac{df}{dx}\right)^{-2} \;,
	\label{sfit2}
\end{eqnarray}
where $df/dx$ is the slope of the curve $y=f(x)$. Note, by definitions, $s_y$ 
is always a constant, but $s_x$ is a function of $x$ unless $f(x)$ is a linear 
function of $x$.

We measure the scatter of data points around the curve $y=f(x)$ by the 
variance of fit along a direction locally perpendicular to the curve, 
which is (see Fig.~\ref{fig4})
\begin{eqnarray}
	s^2 = \frac{s_x^2 s_y^2}{s_x^2 +s_y^2} = \frac{s_y^2}{1+(df/dx)^2}\;.
\end{eqnarray}

For the problem in this paper, we have $x=\log V$, $y=\log L$, and $f(x) = 
a x +b$. Thus, $df/dx = a$ is a constant, and $s_x = s_y/a$, $\sigma_i^2 = 
\sigma_{y,i}^2 + a^2 \sigma_{x,i}^2$. The {\em deviation of fit} is then
\begin{eqnarray}
	s = \frac{s_y}{\sqrt{1+a^2}} \;.
\end{eqnarray}

\begin{figure}
\vspace{2pt}
\includegraphics[angle=0,scale=0.46]{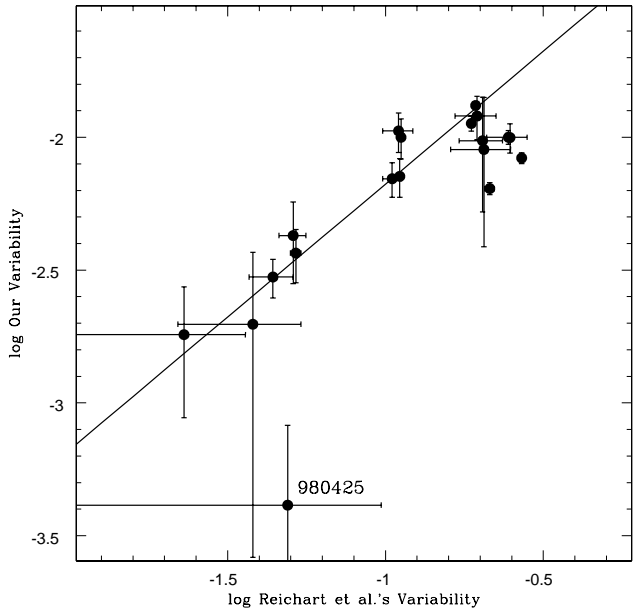}
\caption{The variability with our definition versus the variability with Reichart 
et al.'s definition for 19 GRBs. The straight line is defined by the equation
``our variability $=$ Reichart et al.'s variability divided by 15''.
}
\label{fig5}
\end{figure}

With the above tools, we are ready to compare our results to that of G05
quantitatively. To make the comparison well defined, we analyze the 
overlapping sample
which contains 19 GRBs (see Section~3). In Fig.~\ref{fig5} we plot the 
variability with our definition versus that with R01's definition (the latter
was copied from G05). The two variabilities are highly correlated, but the 
scatter is large. This large scatter is necessary for obtaining a better 
correlation between the peak luminosity and the variability of GRBs. 

The largest scatter in Fig.~\ref{fig5} occurs at GRB 980425, for which our
background fit leads to $T_{90} = 60.29$ sec, in contrast to the $34.88$ sec 
reported in the BATSE web page. If we adopted $T_{90} = 34.88$ and $\xi
=1$ ($\xi$ is the ratio of the background fluctuation to the Poisson noise,
see Section~2), we would obtain $V = 0.00053\pm 0.00057$ for 980425. This
value of $V$ is only slightly larger than that listed in Table~1, so 980425 
remains to have the largest scatter in Fig.~\ref{fig5}. With our definition
of variability, GRB 980425 is much less offset from the variability-luminosity
correlation.

Figure~\ref{fig5} also shows that the values of the variability  with our 
definition are systematically smaller than that with R01's definition, as
indicated by the straight line defined by ``our variability $=$ R01's
variability divided by 15''. This is mainly caused by the different 
normalization in the two definitions. In addition, our variability generally
has a larger error bar than that of R01, caused by the fact that we normalize
our variability by peak count, which suffers larger photon noise than total 
counts; and we have considered the error arising from changing $N_{\rm iter}$ 
by $\pm 1$.

Now, we make a linear fit to $\log L-\log V$ for the overlapping sample, with 
our definition of variability. For the reasons explained in Section~4, we 
exclude GRBs 980425 and 030329 during the fit. Thus, the total number of GRBs
to fit is $N=17$. We obtain $\log L = 3.10 \log V + 59.13$, and $\chi^2/\dof 
= 2.43$ where $\dof = 15$. The weighted average of the individual data 
variance is $\langle \sigma_i^2\rangle =  0.02715$. Thus, for the fit with 
our definition of variability, we have $s_y = 0.257$, $s_x = 0.0829$, and 
$s =0.079$. 

\begin{figure}
\vspace{2pt}
\includegraphics[angle=0,scale=1.07]{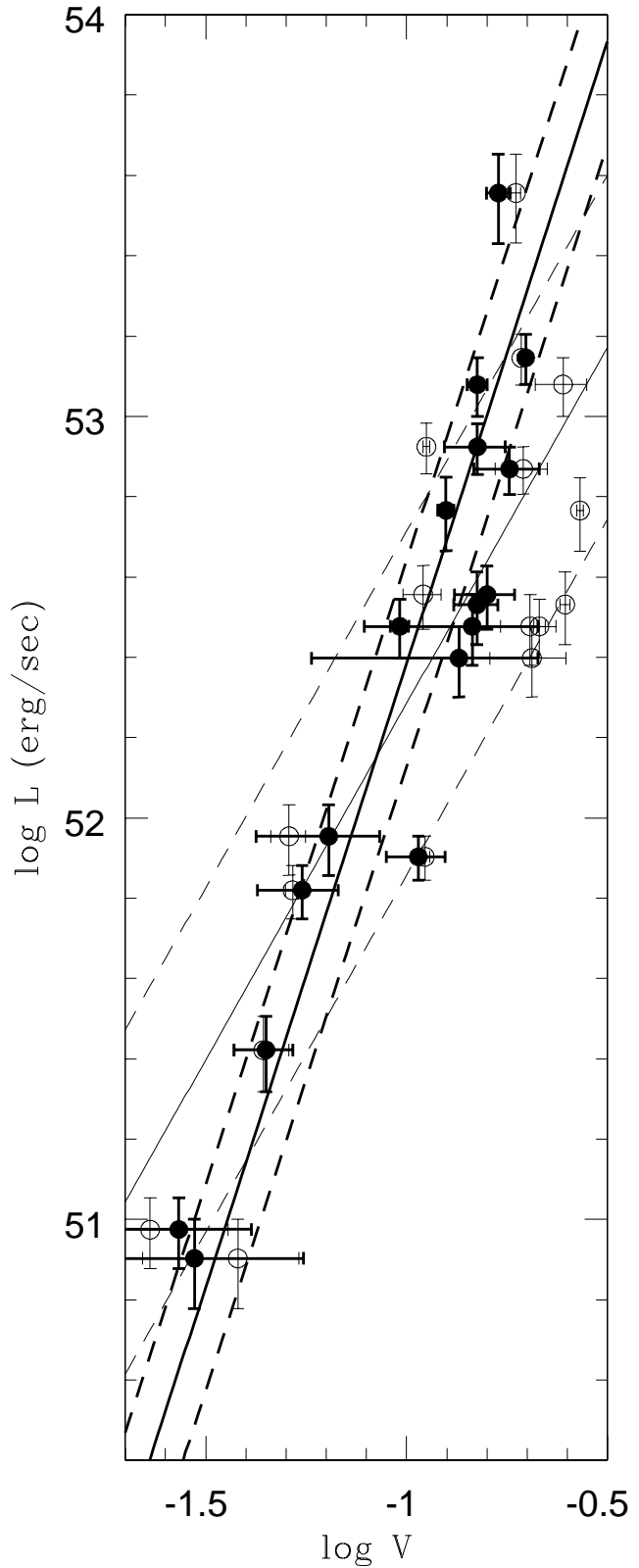}
\caption{Comparison of our results (filled circles and thick lines) to that 
of G05 (open circles and thin lines). Straight lines represent the least-$\chi^2$ 
fit and the 1-$\sigma$ width. (Our variability has been multiplied by a factor
15.)
}
\label{fig6}
\end{figure}

Next, we make a linear fit to $\log L-\log V$ for the overlapping sample, with 
R01's definition of variability. Again, GRBs 980425 and 030329 are excluded,
to make the GRB members in the sample remain the same. We obtain $\log L = 
1.77\log V + 54.06$, and $\chi^2/\dof = 19.89$. The weighted average of the 
individual data variance is $\langle \sigma_i^2\rangle =  0.009240$. Thus, 
for the fit with R01's definition of variability, we have $s_y = 0.429$, 
$s_x = 0.2417$, and $s =0.211$.

The scatter of data points around the fitted $\log L-\log V$ is measured by 
$s$. Thus, according to the above numbers, for the overlapping sample the 
scatter in our $\log L - \log V$ relation is smaller than that of G05 by a 
factor of $0.211/0.079 = 2.7$, although the weighted average of the individual 
data error ($\equiv \langle\sigma_i^2\rangle^{1/2}$) with our definition of 
variability is larger than that with R01's definition of variability by a 
factor of $(0.02715/0.009240)^{1/2} = 1.7$.

The above results, for the overlapping sample excluding 980425 and 030329, are 
summarized in Fig.~\ref{fig6}, where filled circles and thick lines represent 
the results with our data, open circles and thin lines represent the results
with G05's data. The luminosities in the two set of data are the same, so each 
GRB is represented by a horizontal line in Fig.~\ref{fig6}. However, the
variabilities in the two set of data differ, so we see the offset along
the horizontal direction of each pair of filled circle-open circle. The solid
straight lines are least-$\chi^2$ fits to each data set, and the dashed
straight lines mark the ``1-$\sigma$ width'' of the fits. We have set the
unitary length of the horizontal and vertical axes to be the same, so that 
for each data set $s$ is just the half-width of the region bounded by the two
dashed lines that you see. The figure clearly shows that the scatter of our 
data set is much narrower than that of G05's data set.

With the above defined scatter parameter $s$, we can also check how sensitive 
our results are to the parameters in our definition of variability. As 
mentioned in Section~2, we have chosen $T_f = T_{0.5}$, and $N_{\rm iter}$ to 
be the integer closest to the ratio $T_{90}/T_f$. With this choice, for the 
GRB sample described in Section~3 (excluding 980425, 030329, and 030528), we 
obtained $\chi^2/\dof = 1.93$ and $s = 0.078$ (Fig.~\ref{fig1}, where the two 
dashed lines are obtained by shifting the solid line upward/downward by $s_y 
= 0.265$). If we followed R01 and G04 to choose $T_f = T_{0.45}$, we would 
obtain $\chi^2/\dof=2.53$ and $s = 0.087$ in the linear fit to $\log L-\log 
V$. A slightly larger value of $s$ indicates that the correlation is slightly 
looser.

\begin{figure}
\vspace{2pt}
\includegraphics[angle=0,scale=0.45]{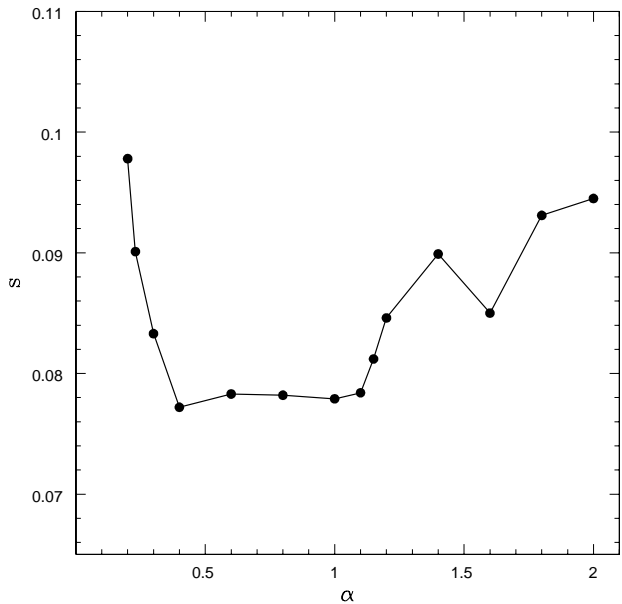}
\caption{Variation of the deviation of the linear fit to $\log L - \log V$, 
with respect to the choice of $N_{\rm iter}$. The parameter $\alpha$, which the 
$T_{0.5}$ in the definition of $N_{\rm iter}$ is multiplied by, is set to be 
the same for all GRBs in the sample. The variabilities presented in this 
paper correspond to the case $\alpha=1$.
}
\label{fig7}
\end{figure}

To test how sensitive the results are to the choice of $N_{\rm iter}$, let us 
redefine $N_{\rm iter}$ by the integer closest to $T_{90}/\alpha T_{0.5}$, 
where $\alpha$ is a parameter that is set to be the same for all GRBs in the 
sample. When $\alpha =1$, we return to the definition that we have adopted. We 
now let $\alpha$ vary from $0.2$ to $2$,  and calculate the corresponding 
$s$ in the least-$\chi^2$ fit to $\log L-\log V$. The calculated $s$ versus 
$\alpha$ are presented in Fig.~\ref{fig7}, which shows that the results are 
stable to the choice of $N_{\rm iter}$. That is, the variance of fit does not 
change drastically with a small change in $\alpha$. The figure also shows that
$\alpha = 1$ is around the minimum of $s(\alpha)$.

\section{Conclusions and Discussion}

We have presented a new definition for the variability of GRB lightcurves. The
new variability is defined in the observer's frame, and the lightcurve is 
smoothed with the Savitzky-Golay filter. The former makes it easy to apply the 
procedure to all GRBs with well measured lightcurves (i.e., with a significant 
signal-to-noise ratio, and a reasonable size of time bin---neither too large 
to erase the variability on short timescales, nor too small to introduce too 
large Poisson noises), without referring to the redshift information.

We have applied our smoothing procedure to a sample of 25 long duration GRBs 
with measured redshifts and publicly available data. A very tight correlation 
between the variability and the peak luminosity is found, with only one 
prominent outlier: GRB 030329. Excluding 030329, 980425, and 030528 (the latter 
two are due to their too large errors in variability), a linear fit to the log 
of variability and the log of peak luminosity is obtained (equations~\ref{lllv} 
and \ref{ab}), with $\chi^2/\dof = 1.93$. The smallness of the reduced 
$\chi^2$ indicates that the data scatter is very small (Fig.~\ref{fig1} and
Fig.~\ref{fig6}). 

We note that, although the existence of a second, larger episode of emission 
from GRB 050820a has been claimed \citep{gol05,pal05,cum05} and we have only 
made use of the data for its first episode, GRB 050820a fits the relation 
perfectly well.

Our results are a significant improvement to that of G05 (as well as that of 
FR00 and R01), who used a different definition of variability and obtained 
$\chi^2/\dof = 1167/30= 38.9$. An analysis on the overlapping sample of
17 GRBs (after GRBs 980425 and 030329 being excluded) shows that the scatter
in our data is smaller than that in G05's data by a factor of $2.7$, measured
by the deviation of fit (Section~5).

The improvement to the correlation is caused by using a new smoothing procedure 
and a different normalization in the definition of our variability, not by 
neglecting the effect of GRB redshift. R01 have shown that the dependence of 
variability on the redshift is extremely weak. The fact that a very tight 
correlation between the peak luminosity and the variability is obtained 
without including the corrections to the variability from the GRB redshift 
also leads us to believe that such corrections are not necessary. 

A remarkable feature of the correlation that we have found is that it does 
not rely on the corrections to the luminosity from the collimation of GRB 
jets. In other words, the peak luminosity used above is simply the 
isotropic-equivalent peak luminosity. If this correlation is confirmed by 
future bursts, it will provide a convenient calibration to GRB luminosity
and distance.

GRB 030329, which is considered to be a firm case for the connection of GRBs 
with supernovae \citep{sta03,hjo03}, is offset from the correlation by 7
$\sigma$ (Fig.~1). Although the cause is not clear, we have noticed that 
030329 and 050525 differ from other GRBs in the sample in the following way: 
their lightcurves consist of two distinctly separated pulses with each
containing at least 30\% of total counts. Motivated by this observation and the
fact that GRB 050820a fits the relation well although only the first episode
of its emission has been used, we have recalculated the variabilities of 030329 
and 050525 by dividing the smoothing window by a factor of 2 (i.e., dividing the 
$n_{\rm P}$ obtained from $T_{0.5}$ by 2) but keeping $N_{\rm iter}$ 
unchanged.\footnote{Roughly speaking, this is equivalent to smooth each GRB
pulse separately.} We obtained $V = 0.0022\pm 0.0003$ 
for 030329, which well fits the solid straight line in Fig.~1. For 050525, we 
obtained $V = 0.0032\pm 0.0007$, which fits the straight line equally well as 
the value listed in Table~1. Whether this treatment is correct must be tested 
when more GRBs with lightcurves similar to that of 030329 and 050525 are 
available.

\section*{Acknowledgments}

LXL thanks C. Guidorzi and T. Sakamoto for useful communications. The 
authors thank the anonymous referee for a very helpful report that has 
led to significant improvements to the paper. 
This research has made use of BATSE and BAT/{\it Swift} data obtained from 
the High-Energy Astrophysics Science Archive Research Center (HEASARC), 
provided by NASA Goddard Space Flight Center; and data obtained from the 
\het\ science team via the website http://space.mit.edu/HETE/Bursts.

\appendix

\section[]{Analytic Approximation to the Error in the log Luminosity}

The numerical results for the error in the $\log L$ predicted by
equation~(\ref{lllv}) due to the uncertainties in $a$ and $b$, $\sigma_{ab}$,
can be approximated by the following formulae ($\sigma_{ab,1}$ for the upward
error, $\sigma_{ab,2}$ for the downward error): 

(1)~When $\log L < 52$
\begin{eqnarray}
	\sigma_{ab,1} &\approx& 7.2017 - 0.16777 x + 0.0006006 x^2 \;,
	\label{eq1}\\
	\sigma_{ab,2} &\approx& 8.0929 - 0.18000 x + 0.0005134 x^2 \;,
	\label{eq2}
\end{eqnarray}
where $x\equiv\log L$, $L$ is in units of erg/sec. (2)~When $52 \le \log L < 
54$
\begin{eqnarray}
	\sigma_{ab,1} &\approx&  2274.034248 - 124.984716 x + 2.286986 x^2 
		\nonumber\\
		&& - 0.013930421 x^3 \;, \label{eq3}\\
	\sigma_{ab,2} &\approx&  3028.186992 - 167.524125 x + 3.087093 x^2 
		\nonumber\\
		&& - 0.018948436 x^3 \;. \label{eq4}
\end{eqnarray}
(3)~When $\log L \ge 54$
\begin{eqnarray}
	\sigma_{ab,1} &\approx& -6.8502 + 0.12989 x \;, 
	\label{eq5} \\
	\sigma_{ab,2} &\approx& -5.7594 + 0.10933 x \;.
	\label{eq6}
\end{eqnarray}

The fractional errors in $\sigma_{ab}$ given by these formulae are always 
$<5\%$.

\bsp

\label{lastpage}

\end{document}